\documentstyle[12pt]{article}
\evensidemargin\oddsidemargin
\begin{document}
\count0 = 1
\begin{titlepage}
\title{\small{ASYMPTOTIC STATE VECTOR COLLAPSE,\\
AND UNITARILY NONEQUIVALENT
REPRESENTATIONS OF QED \\ }}
\author{S.N.Mayburov \thanks{E-mail ~~ Mayburov@SCI.LPI.MSK.SU ~~
 }\\
Lebedev Inst. of Physics\\
Leninsky Prospect 53\\
Moscow, Russia, 117924\\
\\}
\date{ Submitted to Appl. Phys. B}
\maketitle
\begin{abstract}
 The state vector evolution in the interaction of initial measured pure state
 with collective quantum system or the field with a very large number of
 degrees of freedom N is analysed in a nonperturbative QED formalism. As the
 example the measurement of the electron final state scattered on nucleus or
 neutrino is considered.In the nonperturbative field theory (QFT) the complete
 manifold of the system states is nonseparable i.e. is described by  tensor
 product of infinitely many independent Hilbert spaces. The interaction of
 this system with the measured state can result in the final states which
 belong to different Hilbert spaces which corresponds to different values of
 some classical observables,i.e. spontaneous symmetry breaking occurs.
 Interference terms (IT) between such states in the measurement of any
 Hermitian observable are infinitely small and due to it the final pure states
 can't be distinguished from the mixed ones, characteristic for the state
 collapse. The evolution from initial to final system state is nonunitary and
 become formally irreversible in the limit of infinite time. The
 electromagnetic (e-m) bremmstrahlung produced in the electron scattering
 process contain the unrestricted number of soft photons which radiation flux
 become classic observable. Analoguous processes which occurs in the second
 kind phase transitions in ferromagnetic and phonon excitations in cristall
 lattice are considered briefly.
\end{abstract}
\end{titlepage}
1. The problem of the state vector collapse description in Quantum
Mechanics (QM) is still open despite the great efforts. The multiple
 proposals including such exotic as many worlds QM interpretations
,or collapse in the human brain constitute a long list, but in our
 opinion only microscopic dynamical models deserves serious 
 analysis. Of them the most well developed are so called 
 Decoherence models $\cite{Zur}$, despite they have serious
 internal consistency
 problems ,so called Environement Observables Paradox (EOP)  $\cite{Desp}$.
 In its essence it means that for any decoherence process exists at least
 one observable $\hat{B}$ which expectation value will coincide with the
  pure final state expectations and differ from the
 predicted for the mixed state. 
 Meanwhile Neeman proposed in 1985 that the solution of the 
 problem can lay beyond the realm of standard QM
 and must be studied by the methods of modern nonperturbative QFT -    
 dynamiclal theory with infinite number of degrees of freedom $\cite{Nee}$. 
 He considered the analogy between the collapse and the superconductor
  state evolution ,where spontaneous symmetry breaking of initial state
  result in appearence two different Hilbert spaces for quasyparticles
which released with equal probability
 In this approach EOP is resolved in a natural way ,due to the 
 unobservability of interference between states from different 
 spaces. Now this approach needs furhter detailization and 
 application of some calculable models. In this paper we consider
 the model of this in the nonperturbative QED framework.

2. First we remind some nonperturbative QED results.
 In the  perturbative QED framework for the electron
 $P_e\rightarrow P'_e$ scattered  on some target the probability to emit photon
 with energy less than $\hbar\omega$
 is proportional to $e^2ln\frac{P_e}{w}$ and so it grows unrestrictedly as
 $w\rightarrow 0$. In other words QED have running interaction constant
 $\alpha_r(\frac{P_e}{\omega})$ which diverge in this limit.
 It shows that perturbative methods are useless in this region where
 the number of simultaneosly produced photons can be very large .
 The novel nonperturbative formalism which permit to calculate QED S-matrix
 correctly was developed in the last years $\cite{Kib}$.
 It was proposed first for the case when electron evolution in momentum
 space can be regarded classicaly ,i.e. $P_e,P'_e$ are eigenvalues of
 in,out states and electromagnetic current $J_{\mu}(x)$ is not
 the operator .but c-value which Fourier transform is equal to
\begin{equation}  
    J_{\mu}=J_{\mu}(k,p,p')=ie(\frac{p_{\mu}}{pk}-\frac{p'_{\mu}}{p'k})   
\end{equation} 
 This formalism is correct if the recoil of radiated photons is small
 and doesn't change $e$ final momentum, which is true for most cases.  

Final photon state is calculated in the S-matrix approach by the action of    
  operator T-product of interaction Hamiltonian density
$\hat{H}_{em}=J_{\mu}(x)A_{\mu}(x)$on the
 initial vacuum state $\cite{Kib}$
\begin {equation} 
     |f\rangle=|\gamma^f\rangle=S(J)|0\rangle=exp(i\phi(J)-V(J)+U(J))|0\rangle
\end {equation}
where
$$
    V(J)=.5\int {d\tilde{k}J^*(k)J(k)}
$$
$$
 U(J)=i\int d\tilde{k}(J_{\mu}(k)a^+_{\mu}(k)-J^*_{\mu}(k)a_{\mu}(k))
$$
where $d\tilde{k}=\frac{d^3k}{k_0}$.
Here $\phi(J)$ is the pure phase,which is infinite for $J_{\mu}$ of (1) : 
$$
  \phi(J)=\int\frac{d^4kJ(k)J(-k)}{2(2\pi)^4k^2}
$$

This  results in the divergent photon spectra
\begin {equation}   
 d\bar{N}=c\frac{d\omega}{\omega}=\frac{d\tilde{k}}{\hbar}|J(k)|^2
\end {equation}
well known in the classic theory. In the same time we get from (2) 
$$
|\langle f|0\rangle|=exp(-\frac{\bar{N}}{2})=0
$$
This case is equivalent to Bogolubov boson transformation
 of photon free field operators $a_{\mu}(k),a^*_{\mu}(k)$
\begin {equation} 
     b_{\mu}(k)=a_{\mu}(k)+iJ_{\mu}(k)
\end {equation}
 This transformation is nonunitary for $J_{\mu}$ given by (1) and
the obtained final state doesn't belong to initial Fock space $H_F$.
$H_F$ is
 separable Hilbert space,i.e. can contain only states with $ N<N_{max}$
,where $N_{max}$ is arbitrary large . In considered theory 
 complete states manifold is nonseparable ,i.e. described by the tensor
 product of the
infinitely many Hilbert spaces $H_i,$, where $i$ index depends on the
boson transformation constant $J_{\mu}(k) (J_{\mu}(k)=0$ for all $k$
corresponds to $H_F$).    

In the described QED approximation $|f\rangle$ will be  generalised coherent
 state with infinite norm as can be seen from (2) $\cite{Its}$.
This state is the cyclyc vector in $H_j$ and define the new vacuum
state $|0\rangle_j$ from which we can constaruct all other vectors of $H_j$ .  
Any Hermitian operator $\hat{B}$ - observable acts only inside  a single
space $|\psi'\rangle_i=\hat{B}|\psi\rangle_i$ ,so that
 $\langle_i\psi|\hat{B}|g\rangle_j=0$.
So if the final state is the superposition of the states 
from  several spaces $|f\rangle=
|f_1\rangle_i+|f_2\rangle_j$
,any interefrence terms (IT) between $f_1,f_2$ are unobservable and this
states notified as disjoint states  can't be distinguished from the mixture.

3. As the model of the state vector collapse we regard the weak scattering
of the electon and some neutral particle (neutrino) $\nu$ with mass $m_0$
 without assuming any classical properties of final states.

 Amplitude $M_w$ of weak vertex $e,\nu\rightarrow e',\nu'$
 corresponds to finite cross-section and spherically
symmetric distribution of $e'$,$\nu'$ $\cite{Its}$:
\begin {equation} 
   M_w=\frac{G}{\sqrt{2}}J_{L\mu}J^*_{L\mu}=\bar{u_e}\gamma_{\mu}
  (1+\gamma_5)u_{\nu}\bar{u'_e}\gamma_{mu}(1+\gamma_5)u'_{nu}
\end {equation}
Perturbative amplitude of the photon with polarisation $e_{\mu}$ to be radiated
 in this scattering,under condition that $e$ final momentum is $P'$ is:    
\begin {equation} 
      M=M_wM(e\rightarrow e' \gamma)=M_weJ_{\mu}e_{\mu}
\end {equation} 
For general S-matrix calculation defined by $\hat{H}_i=\hat{H}_{em}+\hat{H}_w$
 we use the smallness of weak interaction constant $G$ which permit
to calculate all the weak processes and consequently  $e,\nu$ momentums
 perturbatevly. 
As previously we suppose the spectrum of $P_e'$ is mainly defined by $M_w$
neglecting photon recoil . In this approximation $J_{\mu} $ becomes the
operator of fermion fields, but
it conserve to commute with e-m field operators 
$\hat A_{\mu}(x)$ and in fact will define final e-me field state. The final
 system state is now completely nonclassical and is the 
the entangled product of $e$,$\nu$ and e-m field states
\begin {equation} 
    |f_w\rangle=\sum_{l=0}c_l|f_l\rangle=
\sum_{l=1} c_l|e_l\rangle|\nu_l\rangle|\gamma^f_l\rangle
  +c_0|e\rangle|\nu\rangle|0\rangle      
\end {equation}
where sum over $l$ means integral over final $e,\nu$ momentums $p_l,p_{l\nu}$
,$|\gamma^f_l\rangle=S(J_l)|0\rangle$,$J_l=J(k,p,p_l)$, $c_0$ is the rate of
 noninteracting particles, $c_l$ is proportional to corresponding $M_w$.
This formalism can be obtained in general form from the Low theorem
as we discuss below.
Note that this final state in general isn't coherent,which can have important 
consequences. The phases $\phi(J_l)$ are infinite ,moreover their differences
$\delta_{lm}$ are to be infinite also between  disjoint final states
 $f_l,f_m$ corresponding to different Hilbert spaces.
Due to it in the limit $t= \infty$ this process is formally irreversible
,because such infinite difference doesn't permit to define the
relative phases of disjoint final states which must be operateded
 as mixed ones.  
 T-reflection of such final states and the consequent  rescattering
will result in the state completely different from the initial one.       
We don't discuss this effect at length, because in our calculations
we'll not apply it directly. Note only that for classical case of (2)
T-reflection will restore initial state but with some new arbitrary phase.

We want to measure in this layout if the act of scattering took place or
the particles passed untouched and conserved their initial state. In the same
 time it will be the measurement of particles helicity, because at high energy
$\sigma_L>>\sigma_R$ for weak interactions. 
   As the detector we can consider the single molecule $D_2$ which can
 dissociate
in collision with the radiated photon with $E>E_d$ in 2 atoms $D^*$.
\begin {equation} 
    |f_{md}\rangle=\sum|f'_l\rangle=
|D^*\rangle\sum_{l=1} c'_l|e_l\rangle|\nu_l\rangle|\gamma'_l\rangle
  +c_0|D_2\rangle|\nu\rangle|e\rangle|0\rangle      
    \end {equation}
where
$$
 |\gamma'_l\rangle=\int_{E_d}d\tilde{k}f(k)a_(k)|\gamma^f_l\rangle
$$
$f(k)$ is dissociation amplitude.
In our study  we neglect the final states $|f_l\rangle$
for which all photons have
the energy $E<E_d$ regarding them as the detector inefficiency.
Due to (4) $|\gamma'_l\rangle$ is the vector of the same space
$H_l$ to which $|\gamma^f_l\rangle=|\gamma^f\rangle_l$ belong.It 
follows that for any obsevable $\hat{B}$ :
$$
  \langle e_l,\nu_l,D^*|\langle\gamma'_l|\hat{B}|0\rangle|e,\nu,D_2\rangle=0
$$
  So  D final states are entangled with e-m field disjoit states which destroy
IT.

Note that in practice direct $\hat{B}$ IT observation is impossible
even between single photon $|k\rangle$
and vacuum state as follows from Glauber theory 
of photocounting$\cite{Gla}$. To reveal IT presence the special
premeasurement procedure must be done,  $|k\rangle$
must be reabsorbed by its source S and the interference of
source states for some new observable $B_s$ studied. The situation
become even more complicated for nonunitary evolution when $B_s$ 
 can be nonexisting. It certainly will be so for $|f_{md}\rangle$
states at $t=\infty$ due to dicussed loss of relative phases between
its parts $|f'_l\rangle$. Really if the phase differences $\delta_{lm}$
are infinite for disjoint final e-m field states ,then their reabsorbtion
will mean that phase loss is transferred to $S$ states which become mixed. 
 But we'll  show this impossibility also for the experimental  procedure
 performed at large,
 but finite time in the radiation interference formalism  described in
 $\cite{Its}$.  
We don't consider at all the restoration of $D_2$ state,asuming it
completely reversible and consider the rescattering of the state
$|f_w\rangle$ of (7). We'll regard  gedanken experiment where scattered 
 $e,\nu$ are reflected by some very distant mirrors back to the 
inetraction region where they rescatter again. This formalism is applied  
for charged current of $e$ which passed through $n$ consequent collisions
with simple topological and casual structure. Then to calculate 
final state we can use (7) in  which we take $J^s_l=\sum^{n}J^i_l$.
 But for the currents given by (1) we immediately get that
the main part of it is equal to $J_{\mu}(k,p_{in},p^l_{out})$
. In other words all intermediate steps are unimportant in
agreement with Low theorem $\cite{Its}$, which  shows that infrared photon pole
in any process is defined solely by the current calculated between
asymptotic -in,-out states. 
As follows from (2) $ \langle 0|S(J)|0\rangle$
amplitude of $|0\rangle$ restoration is nonzero  only for $J^s_{\mu}=0$
which means that $e$ in and out momentums  coincide,and from momentum
 conservation the same be true for $\nu$. So we must calculate
the probability $P$ of weak process $i\rightarrow n'_l
\rightarrow i$, where $n'_l$ are all possible intermediate
 states,which we suppose have the same spectra as final states. Its calculation
is simplified by the spherical symmetry of weak scattering (5),
so that neglecting difraction we can omit sum over intermediate states 
 and obtain:
$$   
  P=\frac{\int|M_w(n'_1\rightarrow i|^2do_f}
{\int|M_w(n'_1\rightarrow n_o|^2do_f}=0
$$
where where $n'_1$ is arbitrary intermediate state,  $n_o$ is sum over
 final states.  $o_f$ is phase space of final $e$ states which
is isomorfic to spherical surface with $r=1$ with nearly constant
density of final states on it. Then restoration of
initial state corresponds to a single point $r_i$ on this surface. Each
 inifinetely close point to $r_i$ corresponds to another Hilbert space
 generating infinite number of soft photons. So the zero probability
of initial state restoration obtains simple geometrical interpretation
in which  $r_i$ is singular point in phase space which we can omit
without changing any physical result.  
   
The same effects can be expected for the coulomb $e$ scattering
for which bremstrahlung were most often studied, but
due to long range of this force the analysis will be more intricate.

4. We've shown that final states of $e-\nu$ scattering asymptotically
reveal the properties of the mixed state i.e. perform the collapse.
This doen't seems a surprise ,because the classisal features of
electron bremstrahlung states were stressed often ,but to our knowledge
the final $e$ states interference never was analysed. $\cite{Kib}$. 
In the alternative approach developed by Buchholz
this classical properties  results from electic field flow
 conservation constituting additional  superselection rule $\cite{Buch}$.
  This rule based on Gauss law assume that 
Lorentz symmetry for the electron is spontaneously broken. 
 It suppose that the collapse is induced primarily
not by the huge number of radiated photons but the long range
properties of vector potentials which doesn't permit 
the superpositions of charged states with different velocities
 at the infinite time limit. It must result in some  complications    
 in the description of localised charged states structure and
 evolution which will be analised elsewhere.

The real detectors are localised solid objects to which this formalism
doesn't applicable directly. But the general QFT analysis was very succesful
for the solid state phenomena description ,and so we'll scatch here
the possible framework for the collapse models. The simple model
in which collapse is induced by the by the 2nd order phase transition
in ferromagnetic was given in $\cite{May}$ and we'll descuss here its possible
 developments. It's well known that the transition 
from the individual particles (atoms) to quasyparticles - phonons,
magnons in the infinite media can be  described as boson transformation
analoguous to (4) $\cite{Ume}$. The resulting quasiparticles are massless
and the excitation spectra have no gap i.e. infrared divergent. Despite
the media is electrically neutral this quanta readily interact with e-m field,
so any excitation of this system in the vacuum  can be relaxated by the
 infrared divergent radiation (4).

 This idea is also applicable for the finite
system if its surface is regular and transparent for radiation. This
surface can be regarded as the topological defect with 
infinite number of degrees of freedom  which result in
a special kind of boson condensation in its volume$\cite{Umu}$. So 
the system states manifold can be unitarily nonequivalent and the resulting
 quasyparticles 
spectra is infrared divergent. The same we can expect for the
relaxation photons radiated through the crystall surface. We consider
 as the detector the idealised crystall which can be excited by the
high energy neutral or charged particle. We take the initial state
 $|i>$ to be the superposition of 2 localised states with the 
trajectories $x_1,x_2$ passing through and beyond the crystall,
which has to be measured. So  analoguos to (7) the final state
is the superposition of initial vacuum and some new excited state.  
The state $|x_1>$ can kick out or shift several atoms producing
lattice dislocation and excitation. So its final state after
lattice relaxation will include infinite number of soft photons
which constitute the new vacuum completely ortogonal to initial one.

\end{document}